%% file: FeGe_OurArticle.tex
\documentclass[review]{elsarticle}

\usepackage{hyperref}

\journal{Journal of Magnetism and Magnetic Materials}

\begin{document}

\begin{frontmatter}

\title{First-principles study of magnetic interactions in FeGe}

\author[mymainaddress]{Ilya V. Kashin\corref{mycorrespondingauthor}}
\cortext[mycorrespondingauthor]{Corresponding author}
\ead{i.v.kashin@urfu.ru}

\author[mymainaddress]{Sergey N. Andreev}
\author[mymainaddress]{Vladimir V. Mazurenko}

\address[mymainaddress]{Theoretical Physics and Applied Mathematics Department, Ural Federal University, Mira Street 19, 620002 Ekaterinburg, Russia}

\newcommand{\bm}{\bf}

\begin{abstract}
We theoretically study the magnetic properties of iron germanium, known as one of canonical helimagnets. For this purpose we use the real-space spin Hamiltonian and micromagnetic model, derived in terms of Andersen's ``local force theorem'', to describe the low-lying magnetic excitations via spin-polarized Green's function, obtained from the first-principles calculations. The model was designed to numerically evaluate the spin stiffness constant in reciprocal space, in order for  assessment of the contributing itinerant mechanisms. The calculated pairwise exchange interactions reveal the essentiality of Ruderman-Kittel-Kasuya-Yoshida coupling in FeGe. Thus provided mean-field estimation of magnetic transition temperature agrees good with the experimental measurement, underlining the necessity of the comprehensive real/reciprocal space-based approach for a proper description of magnetic excitations in FeGe.
\end{abstract}


\end{frontmatter}


\input{Body}


\bibliography{mybibfile}

\end{document}

%% file: Body.tex
\newcommand{\bm}{\bf}

\section{Introduction}

To date, the experimental observation of collective magnetic excitations like topological spin structures, referred to as skyrmions, in a number of magnets \cite{Skyrmions_Exp1, Skyrmions_Exp2, Skyrmions_Exp3} raises an interest of researchers due to its multiferroic properties and topological Hall effect \cite{Multifer_TopHallEff}. Iron germanium, helimagnet possessed by formation of a skyrmion crystal \cite{Yu_FeGe}, motivated this study by large prospects of application in state-of-the-art memory devices \cite{ModMemDevices}.

Naturally first features of magnetic interplay in metals, which were quantitatively and qualitatively explained in theoretical framework, manifest itself as originated from low-lying excitations near the ground state, commonly of the spin-wave type. In case of its long periodicity the spin-wave dispersion spectrum $E({\bm q})$ could be adjusted by a single parameter, referred to as spin stiffness constant ${\cal D}$. This parameter, reflecting the curvature of $E({\bm q})$ at ${\bm q} \rightarrow 0$, is observable in magnetisation \cite{SpSt_Experiment_MagMeas} and neutron-scattering \cite{SpSt_Experiment_NeuScat_1, SpSt_Experiment_NeuScat_2} experimental measurements.

The widely used approach to numerically estimate ${\cal D}$ is based on the composition of pairwise exchange interactions $J_{ij}$ in real space \cite{Lichtenstein_SpinSpirals_and_D, InfsSpRot}. We should note, that one can reproduce $E({\bm q})$ directly from $J_{ij}$ using Fourier transformation \cite{Bruno}, allowing to describe corresponding spin dynamics by means of effective Hamiltonians \cite{SpinDynamics_1, SpinDynamics_2, SpinDynamics_3}. But if we deal with metals, where Ruderman-Kittel-Kasuya-Yoshida (RKKY) character of interactions can be predominant, there is no unambiguous way to determine the reliable borders of the spatial composition. Indeed, in the work of Pajda \textit{et al.} \cite{Bruno} authors construct the effective classical Heisenberg model by mapping initially itinerant electron system, described by \textit{ab-initio} calculations. This mapping makes necessary to include the additional damping parameter, providing the numerical convergence of the scheme in real space consideration. Kvashnin \textit{et al.} \cite{Kvashnin_about_Fe} showed that in ferromagnetic bcc Fe magnetic orbitals of different symmetry are responsible for Heisenberg and non-Heisenberg type of exchange interactions, underlining the complex character of magnetism in metals. Furthermore, there are local and non-local correlation effects, influencing the magnetic picture and studied using dynamical mean-field theory approaches \cite{Fe_DMFT, Fe_DMFT_SFmod}. 

Previous theoretical works on FeGe were mainly concentrated on first-principles definition of the parameters of the micromagnetic models. Thus, in  Ref. \cite{Blugel_MagneticModel} the spin stiffness constant ${\cal D}$ was evaluated on the basis of the first-principles calculations as 500~meV~$\cdot$~\AA$^2$, which is close to the estimation, carried out in Ref. \cite{Arita} (482 meV~$\cdot$~\AA$^2$) as proportional to the magnetic transition temperature $T_{C}$ = 278 K \cite{FeGe_Exp_Tc}.
At the same time, a real-space Heisenberg Hamiltonian description \cite{Exch_in_FeGe_AIP, FeGe_SpinModels} of FeGe magnetism can also be found appropriate. Therefore, there is a problem of theoretical treatment of Fe($3d$) states magnetism that can be based on a local moments description \cite{Exch_in_FeGe_AIP, FeGe_SpinModels} or itinerant picture \cite{FeGe_Itinerant_1, FeGe_Itinerant_2}. In this situation comprehensive numerical approaches can be more constructive in investigation of the long-periodical helical excitations as possessed by the system.

In present study we derive a micromagnetic model, following Ref. \cite{Lichtenstein_SpinSpirals_and_D}, devoted to investigate the spin-wave excitations in pristine ferromagnetic metals by means of single-site scattering matrix. In order to avoid the aforementioned problems of real space consideration we firstly note that, owing to Andersen's ``local force theorem'' \cite{Mackintosh, Methfessel}, approximating the change of the system's total energy by the corresponding change of one-particle energies, expression for ${\cal D}$ is initially derived in reciprocal space \cite{Lichtenstein_SpinSpirals_and_D}, with no spatial sum required. This expression determines ${\cal D}$ from the spin-polarized Green's function, which is available from DFT calculations, and its derivative over components of the ${\bm k}$ vector. The way of calculation of the latter is the subject of discussion in this work, since the grid size, which is usually enough to provide convergence for all necessary numerical data, appeared completely insufficient for suggested scheme.

To make an analysis of the role that itinerant mechanisms play in formation of helical excitations in FeGe we fulfil its real space treatment by calculating the exchange interactions up to sixth coordination sphere, abridged only by computational limit. The physical validity of thus mounted picture is to be checked by estimation of magnetic transition temperature as proportional to total exchange field of a single magnetic atom. Proposed two-sided approach is expected to support other studies \cite{Blugel_FeGeStudy_1, Blugel_FeGeStudy_2, Blugel_FeGeStudy_3} in investigation of the processes, underlying the formation of collective magnetic excitations in metallic systems.

\section{Method} \label{Sec:Method}
\newcommand{\ud}{\mathrm{d}}
\renewcommand{\Im}{\mathrm{Im}}
\newcommand{\TrL}{\mathrm{Tr_{L}}}
\newcommand{\TrLS}{\mathrm{Tr_{LS}}}
\newcommand{\ttwog}{\mathrm{T_{2g}}}
\newcommand{\eg}{\mathrm{E_{g}}}
\newcommand{\kmesh}{${\bm k}$-mesh }
\renewcommand{\i}{\mathrm{i}}

In order to consider the low-lying magnetic excitations in FeGe, which correspond to ${\bm q} \rightarrow 0$ regime, we use the micromagnetic model, describing the spin-wave dispersion spectrum as $E({\bm q}) = {\cal D} \, q^{2}$ \cite{Bruno}, where $q = |{{\bm q}}|$. In this case, by means of ''local force theorem'', the second variation of system's total energy with respect to mentioned excitations is given by 
\begin{eqnarray}
\delta^{2} E =  -\frac{1}{\pi} \, \int_{-\infty}^{E_{F}} \ud \omega \, \Im \, \TrLS \big[ \delta^{2} H \, G \, + \,
\delta H \, G \, \delta H \, G \big],
\label{SecondVariation_through_GH}
\end{eqnarray}
where $H$ is the Hamiltonian of the system, taken from DFT calculations, $G(\omega) = (\omega - H)^{-1}$ is the Green's function, $E_{F}$ is the Fermi level, $\TrLS$ denotes the trace over orbital (L) and spin (S) indices.

If we introduce the small spin deviations from the ground state of the system, which corresponds to long-periodical spin wave and causes spin density perturbation $\delta M \sim q^{2}$, the spin stiffness constant in terms of phenomenological theory \cite{PhenomenologicalTheory_1, PhenomenologicalTheory_2} can be expressed via $\delta^{2} E$ as
\begin{eqnarray}
{\cal D} = \frac{2 \mu_{B}}{3 M} \, \sum_{\alpha} \frac{\partial^{2} E}{\partial q_{\alpha}^2},
\end{eqnarray}
where $M$ is the magnetic moment per atom and $\mu_{B}$ is the Bohr magneton. Thus, the final expression for the spin stiffness constant is given by \cite{Lichtenstein_SpinSpirals_and_D, InfsSpRot}
\begin{eqnarray}
{\cal D} =
\frac{\mu_{B}}{3 M}
\frac{\Delta^2}{2 \pi}
\int_{-\infty}^{E_{F}} \ud \omega \, 
\Im \, \TrL \bigg[ \sum_{{\bm k} \alpha} \, 
\frac{\partial G^{\uparrow}_{{\bm k}}}{\partial k_{\alpha}}
\frac{\partial G^{\downarrow}_{{\bm k}}}{\partial k_{\alpha}}
\bigg],
\label{D_final_expression}
\end{eqnarray}
where $\Delta$ is the intraatomic exchange splitting.
In order to evaluate the derivatives of Eq. \ref{D_final_expression} firstly we note that
\begin{eqnarray}
\frac{\partial G^{\sigma}_{\bm{k}}}{\partial k_{\alpha}} = - G^{\sigma}_{\bm{k}} \, \frac{\partial H^{\sigma}_{\bm{k}}}{\partial k_{\alpha}} \, G^{\sigma}_{\bm{k}}.
\end{eqnarray}
Try of presenting analytically 
\begin{eqnarray}
\frac{\partial H^{\sigma}_{\bm{k}}}{\partial k_{\alpha}} = \sum_{ij} t^{\sigma}_{ij} \, (\i R^{\alpha}_{ij}) \, \mathrm{exp}(\i \bm{k} \bm{R_{ij}})
\end{eqnarray}
($t^{\sigma}_{ij}$ is the hopping integral, defined in Wannier functions \cite{Wannier} basis) appears fruitless, because it faces the same trouble of real space sum convergence, even exacerbated by $R^{\alpha}_{ij}$ as the multiplier. Therefore, in this study we consider the derivative in terms of finite difference method:
\begin{eqnarray}
\frac{\partial H^{\sigma}_{\bm{k}}}{\partial k_{\alpha}} \approx \frac{1}{2 h_{\alpha}} 
\big(
H^{\sigma}_{k_{\alpha} + h_{\alpha}} - H^{\sigma}_{k_{\alpha} - h_{\alpha}}
\big),
\end{eqnarray}
where $h_{\alpha}$ is the step of the grid.

In order to estimate the relative role of short-range and long-range magnetic interplay in FeGe we calculate the pairwise exchange interactions, which in the frame of infinitesimal spin rotations approximation \cite{InfsSpRot} can be expressed via interatomic Green's function as
\begin{eqnarray}
\label{Exch_InfSpRot}
J_{0j} = \frac{\Delta^2}{4 \pi} \, \int_{-\infty}^{E_{F}} \ud \omega \, \Im \, \TrL \big[ G^{\uparrow}_{0j} \, G^{\downarrow}_{j0} \big],
\end{eqnarray}
where index $0$ denotes the arbitrary chosen ''centered'' Fe atom. Reproducing in that way the total exchange surrounding $J_{0} = \sum_{j} J_{0j}$ in terms of mean-field \cite{PhysRevB.25.5766} yields the magnetic transition temperature estimation as 
\begin{eqnarray}
k_{B} T_{C} = 2J_{0} / 3,
\label{MTrTemp}
\end{eqnarray}
where $k_{B}$ is the Boltzmann constant. Another type of $J_{0j}$ combination allows one to calculate the spin stiffness constant in real space \cite{Lichtenstein_SpinSpirals_and_D, InfsSpRot, Bruno, Lichtenstein_SpSt_via_J}:
\begin{eqnarray}
{\cal D} = \frac{2 \mu_{B}}{3 M} \sum_{j} J_{0j} \, R_{0j}^{2},
\label{SpinStiffness_via_J}
\end{eqnarray}
where $R_{0j} = |{\bm R}_{0j}|$ is the corresponding radius vector. Indeed, because of the variety \cite{InfsSpRot, PhysRevB.25.5766, Exch_in_FeGe_AIP} of methods, which make $J_{0j}$ available to compute, employing Eq. \ref{SpinStiffness_via_J} seems obviously more practically preferable for ${\cal D}$. However, long-range interplay turns the quantitative reliability of this approach for helimagnets questionable, but nevertheless thus obtained ${\cal D}$ can be involved into analysis concerning the essence of itinerant scenario for magnetism.

\section{Results and discussion} \label{Sec:FeGe}

Iron germanium crystallizes in P2$_1$3 structure ($a$ = 4.700 \AA) that lacks inversion symmetry, originating non-negligible Dzyaloshinskii-Moriya interaction. The positions of iron atoms (in cell units) are 
Fe$_{1}~(x,~x,~x)$; Fe$_{2}~(\frac{1}{2}~+~x,~\frac{1}{2}~-~x,~x)$; Fe$_{3}~(x,~\frac{1}{2}~+~x,~\frac{1}{2}~-~x)$; Fe$_{4}~(\frac{1}{2}~-~x,~x,~\frac{1}{2}~+~x)$, where $x$ = 0.137. Experimentally confirmed \cite{Lebech} long periodicity of helimagnetic excitations ($\lambda$~$\approx$~700~\AA, about 149 unit cells) indicates infinitesimal spin rotations regime valid. The magnetic transition temperature $T_{C}$ = 278 K is close to the room one, making FeGe the promising candidate for many technological applications \cite{Yu_FeGe, ModMemDevices}.

\begin{figure}[h]
\centering
\includegraphics[width=0.8\textwidth,angle=0]{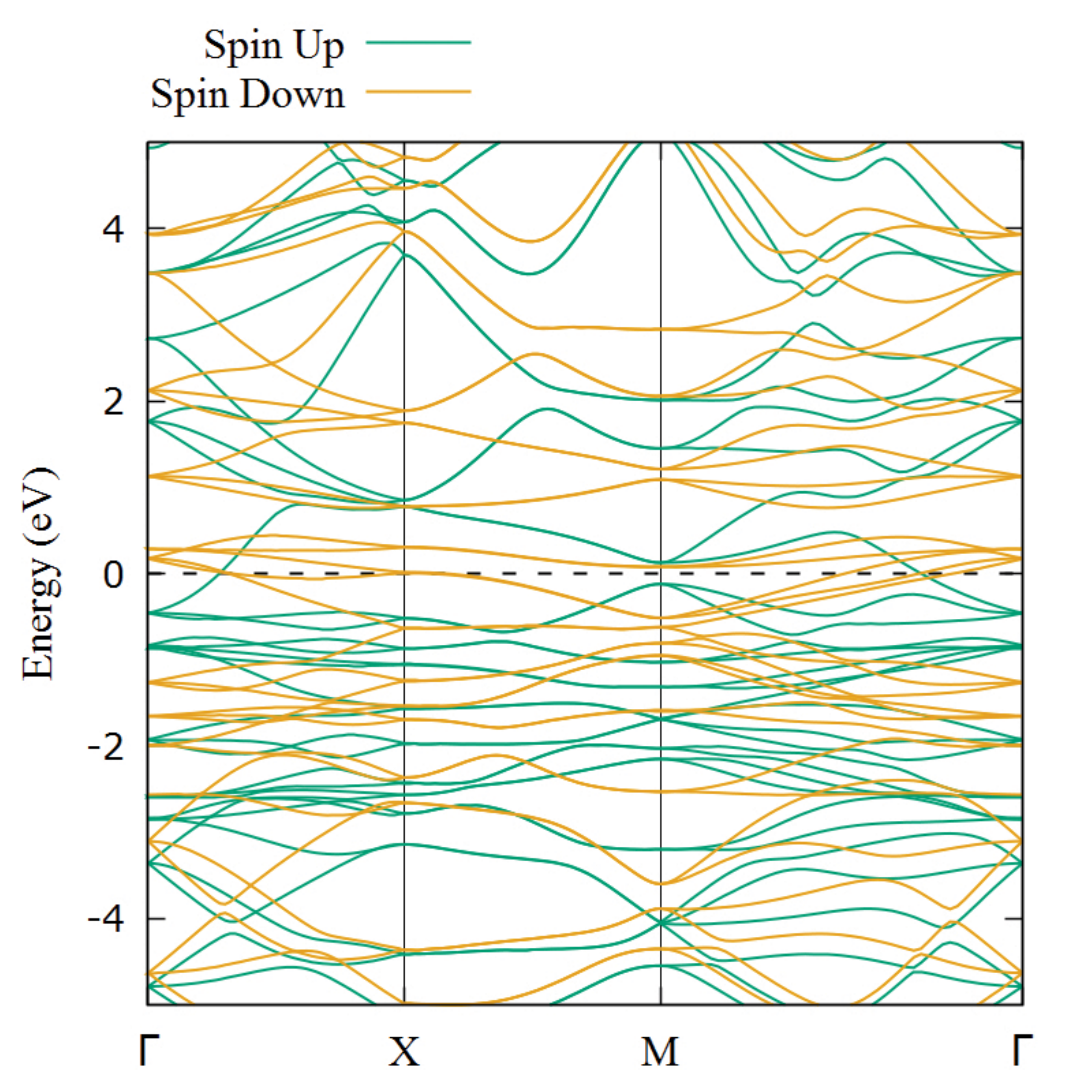}
\caption{The band structure of FeGe, obtained from DFT calculations. The high symmetry points are $\Gamma$(0, 0, 0), X(0,$\frac{1}{2}$, 0), M($\frac{1}{2}$,$\frac{1}{2}$, 0). The Fermi level is shown by dashed line. \textit{(color online)}}
\label{FeGe_Bands}
\end{figure}

\paragraph{DFT calculations}
As the starting point we performed the first-principles calculations of FeGe electronic structure within density functional theory \cite{dft} (DFT) using the generalized gradient approximation (GGA) with the Perdew-Burke-Ernzerhof (PBE) exchange-correlation functional \cite{pbe}. For this purpose the Quantum-Espresso simulation package \cite{QuantumEspresso} was used. The basic parameters of the simulation are following: the energy cutoff of the plane wave basis construction is set to 320 eV; the energy convergence criterion is 10$^{-8}$ eV; the 20$\times$20$\times$20 Monkhorst-Pack mesh was employed to carry out integration over the Brillouin zone. Fig. \ref{FeGe_Bands} gives the resulting band structure, which is in excellent agreement with that presented in previous theoretical studies \cite{Arita, FeGeBands}.

\begin{figure}[h]
\centering
\includegraphics[width=0.8\textwidth,angle=0]{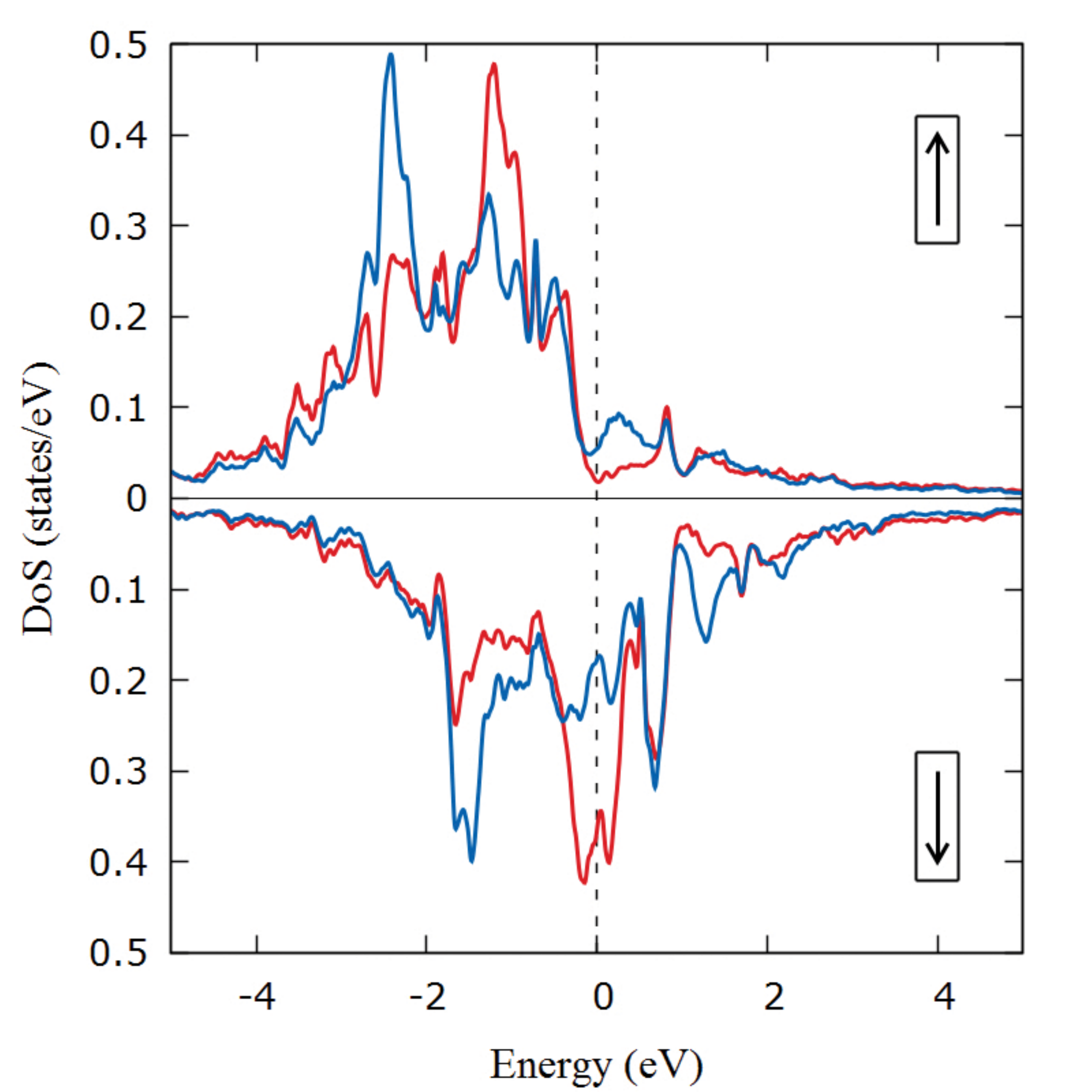}
\caption{Partial density of states of Fe($3d$) shell obtained from low-energy model. Red and blue line denotes $\eg$ and $\ttwog$ states, respectively. The Fermi level is shown by dashed line. \textit{(color online)}}
\label{PartialFeDoS}
\end{figure}

To construct the low-energy model of FeGe magnetoactive electron shell we projected the GGA wave functions onto maximally localized Wannier functions \cite{wannier90, wannier901, wannier902}, describing along with the Fe($3d$) states also Fe($4s, 4p$) and Ge($3d, 4s, 4p$) states, because of the essential entanglement of corresponding bands. The partial density of states of Fe($3d$) shell is illustrated on Fig. \ref{PartialFeDoS}. In contrast to bcc Fe \cite{Kvashnin_about_Fe} for which $3d$ orbitals of $E_{g}$ symmetry demonstrate a localized behaviour on the level of the density of states, we observe a strong delocalization for both $T_{2g}$ and $E_{g}$ states in the case of FeGe. The latter is a result of the strong hybridization between 3d states of iron and 4p states of germanium. As we will show below, the delocalization of the magnetic moments provides a RKKY type of the magnetic couplings in FeGe. 

The intraatomic exchange splitting in Fe($3d$) shell of FeGe is 1.17~eV. The local magnetic moment is 1.18~$\mu_{B}$ per each Fe atom, which is confirmed by previous calculations \cite{Arita} and consistent with the experimental data \cite{FeGe_Experiment_1, FeGe_Experiment_2}. It could raise the idea to attribute FeGe to the case of S = 1/2 system. However, the obstacle is aforementioned intense hybridization, making magnetic moment strongly delocalized and, therefore, itinerant magnetism contribution essential. 

\paragraph{Real space consideration}
We start theoretical investigation of the low-lying magnetic excitations in FeGe by exchange interactions evaluation, according to Eq. \ref{Exch_InfSpRot}. Thus mounted magnetic surrounding was limited by sixth coordination sphere to keep the $\mathrm{exp}(\i \bm{k} \bm{R_{ij}})$ factor numerically stable while calculating the corresponding interatomic Green's functions. As it is seen from Table \ref{ExchangeInt}, our values are in a reasonable agreement with that, calculated by constrained moment approach \cite{Constrained_moment_method} up to third nearest neighbours \cite{Exch_in_FeGe_AIP}. However, application of this approach is conjuncted with explicit simulation of the supercell, making the long-periodical excitations troublesome to capture. Our approach, being much less limited by available computational resources, clearly demonstrate the oscillating RKKY character of magnetism, revealing significance of long-range interactions in FeGe, as well as it was stated for bcc Fe \cite{Kvashnin_about_Fe}.

\begin{table}[h]
\centering
\caption [Bset]{Calculated values of isotropic pairwise exchange interactions $J_{0j}$ (in meV) between magnetic moments of iron atoms in FeGe. CS labels the sequential number of coordination sphere, and ${\bm r}_{0j}$ illustrates the symmetry of corresponding radius vector (in cell units). Each coordination sphere contains 6 Fe atoms. $J^{CM}_{0j}$ (in meV) denotes the values, obtained using the constrained moment calculations \cite{Exch_in_FeGe_AIP}. Note that all exchange interactions are presented as of a single pair of atoms.}
\label {ExchangeInt}
\setlength{\tabcolsep}{7pt}
\begin {tabular}{cccc}
 \hline  \hline
 CS &  ${\bm r}_{0j}$  &   $J_{0j}$  & $J^{CM}_{0j}$ \\
 \hline 
   I  &  $( 0.5,  0.23, -0.27 )$  &    $4.87$  &    $6.35$  \\
  II  &  $( 0.5,  0.23,  0.73 )$  &   $-0.69$  &   $-0.81$  \\
 III  &  $( 0.5, -0.77, -0.27 )$  &   $-0.89$  &   $-2.76$  \\
  IV  &  $( 1, 0, 0 )$            &    $1.25$  &     ---    \\
   V  &  $( 0.5, -0.77,  0.73 )$  &    $0.09$  &     ---    \\
  VI  &  $( 0.5,  1.23, -0.27 )$  &    $0.69$  &     ---    \\
 \hline  \hline
\end {tabular}
\end {table}

Employing Eq. \ref{MTrTemp} yields $T_{C} = 232~K$, which reproduces the experimental measurement reasonably well. It supports the general validity of real space approach to qualitatively explain the basic magnetic properties of helimagnets.

\begin{figure}[h]
\centering
\includegraphics[width=0.8\textwidth,angle=0]{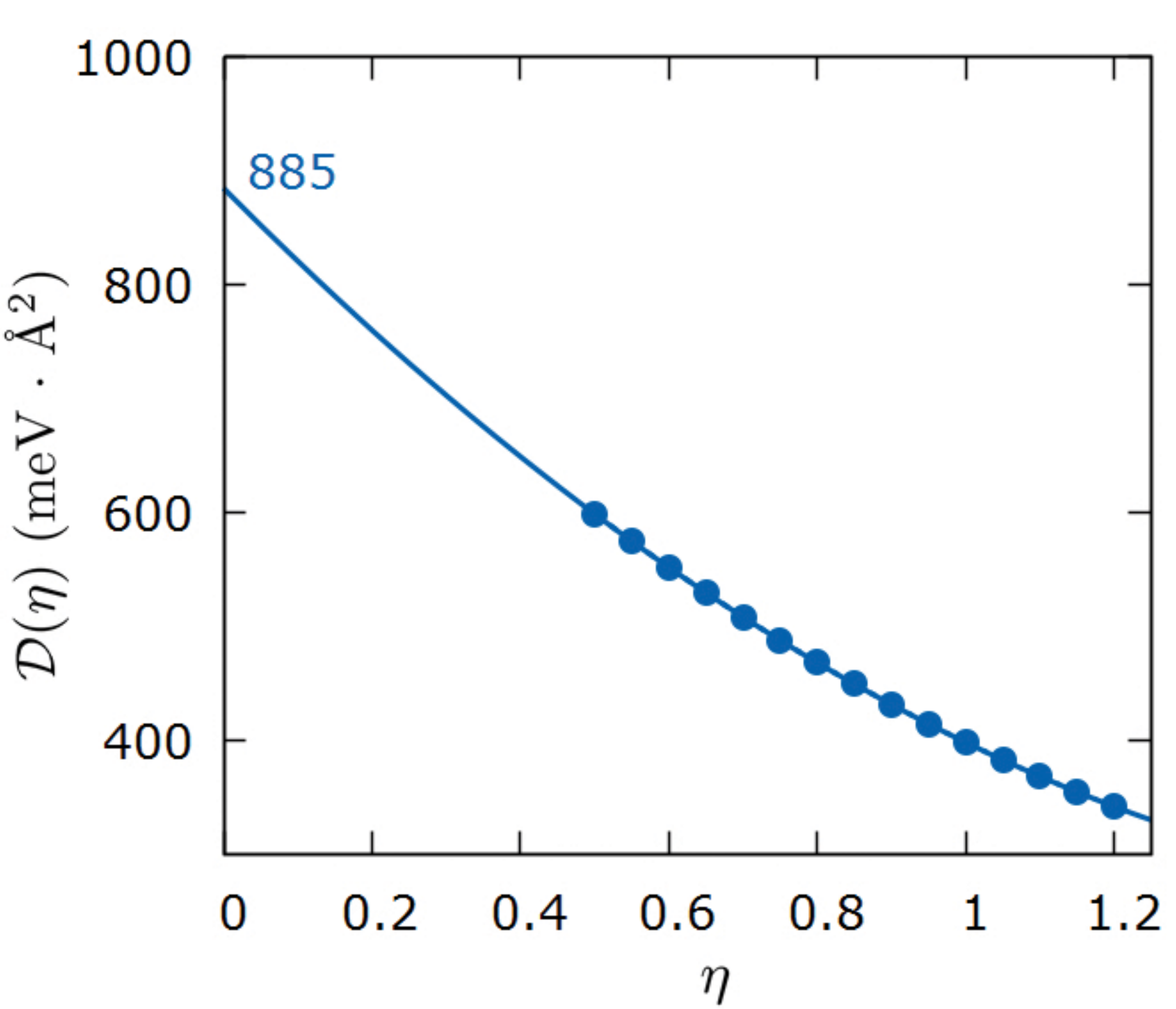}
\caption{Calculated spin-stiffness constant depending on the damping parameter $\eta$. \textit{(color online)}}
\label{D_damping}
\end{figure}

Our discussion of the spin stiffness constant we begin from an important note, that to make a correspondence of the theoretical estimation (both in real and reciprocal spaces) to the experimental value one has to consider it as of the unit cell, destined to keep the symmetry of the system. In FeGe the cell contains four magnetic Fe atoms, successively treated as the ''centered'' one in real space. Thereby, ${\cal D}$ obtained in accordance with Eq. \ref{SpinStiffness_via_J}, turns overestimated (922~meV~$\cdot$~\AA$^2$) in comparison with the experimental evaluation \cite{Arita}. It could be explained by non-localized magnetism as the obstacle for determining the maximum distance, limiting the spatial sum. To inspect the quality of capturing the itinerant aspects of magnetism in such approach we employ the method, proposed by Pajda \textit{et al.} \cite{Bruno} for bcc Fe and applied to Heusler compounds \cite{Heusler_comp_Damping}. This method implies the modification of Eq. \ref{SpinStiffness_via_J} by adding the damping parameter $\eta$, providing the numerical convergence:
\begin{eqnarray}
{\cal D}(\eta) = \frac{2 \mu_{B}}{3 M} \sum_{j} J_{0j} \, R_{0j}^{2} \, \mathrm{exp}(-\eta R_{0j} / a),
\label{SpinStiffness_via_J_with_Damping}
\end{eqnarray}
where $a$ is the lattice constant. Having ${\cal D}(\eta)$ calculated in the range $\eta \in [0.5, 1.2]$, we obtained ${\cal D}(\eta = 0)$ from a quadratic fit as 885 meV~$\cdot$~\AA$^2$ (see Fig. \ref{D_damping}). Only slight improvement of initial overestimation reveals poor capability of the real space approach to describe the helimagnetic state in FeGe, strongly driven by itinerant mechanisms.

\paragraph{Reciprocal space consideration}
At the beginning of reciprocal space consideration we performed the calculations of Eq. \ref{D_final_expression} using the uniform \kmesh to carry the numerical differentiation out. 
The results, shown in Table \ref{D_with_uniform_KMesh}, demonstrate the explicit absence of convergence up to the technical limit of \kmesh size 20$\times$20$\times$20. It can be readily explained by the significant role of itinerant magnetism in FeGe, from one hand, and grid being too sparse for numerical convergence of finite difference method, from another.

\begin{figure}[h]
\centering
\includegraphics[width=0.8\textwidth,angle=0]{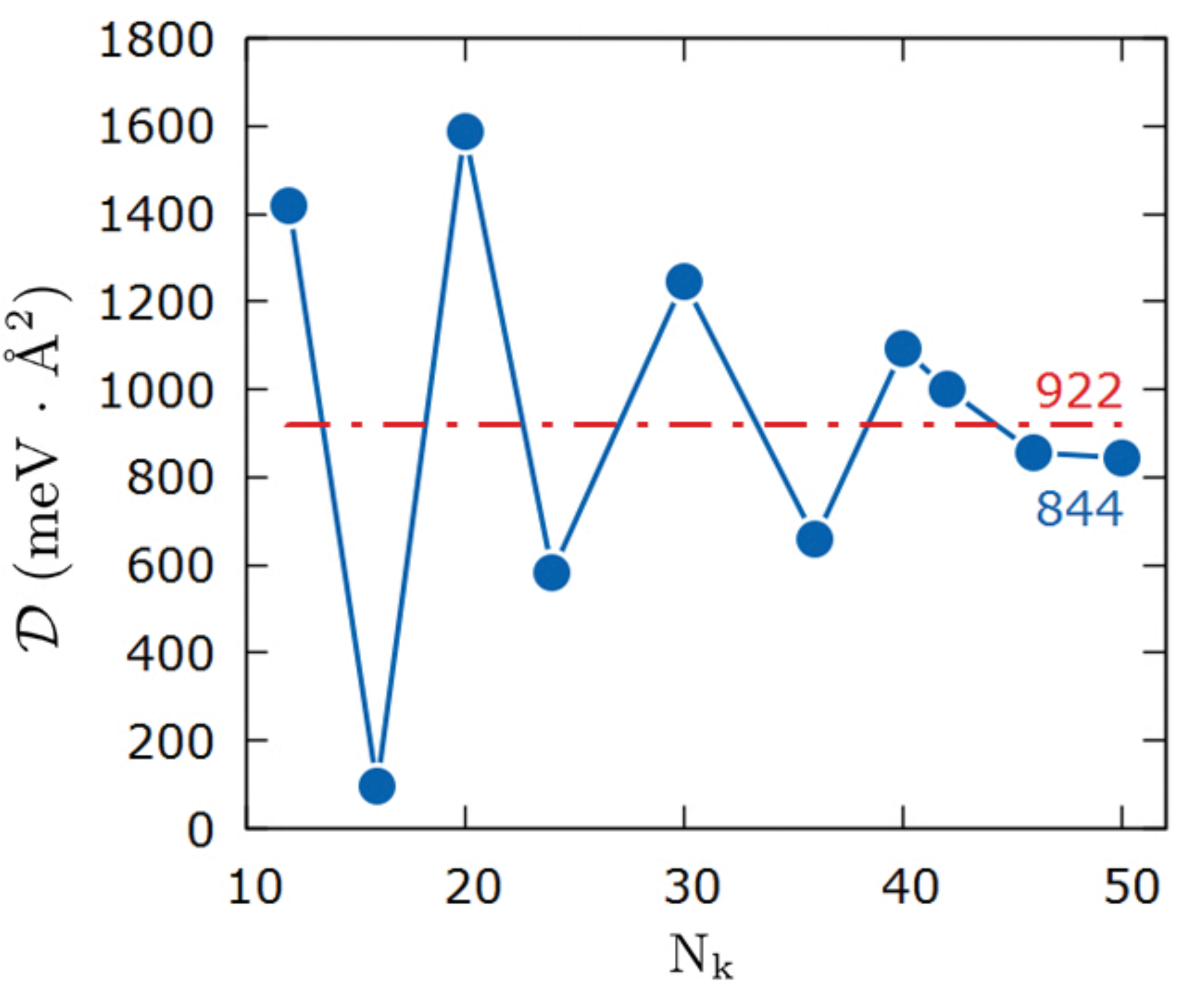}
\caption{The spin stiffness constant as obtained using distinct numerical scheme with non-uniform \kmesh of N$_{\mathrm{k}}$$\times$12$\times$12 size (blue line). Red dot-dashed line denotes ${\cal D}$, calculated in the framework of real space consideration (Eq. \ref{SpinStiffness_via_J}). \textit{(color online)}}
\label{SpinStiffness_N1212_Convergence}
\end{figure}

To circumvent the mentioned limit we set \kmesh to be non-uniform by means of its expanding in one single geometrical direction at expense of other two. From computational point of view, non-uniform grids are known to be widely used tool for solving different numerical problems \cite{FinDiffMethod_NonUnifGrid_1, FinDiffMethod_NonUnifGrid_2, FinDiffMethod_NonUnifGrid_3}. The physical validity of this approach in our case is provided by the diagonal form of the spin-stiffness tensor for cubic systems, engaging in Eq. \ref{D_final_expression} only derivatives over the same geometrical components of ${\bm k}$. The \kmesh configurations N$_{\mathrm{k}}$$\times$12$\times$12, 12$\times$N$_{\mathrm{k}}$$\times$12 and 12$\times$12$\times$N$_{\mathrm{k}}$ were checked to produce the equal values of ${\cal D}$, according to the symmetry. The convergence dynamics is presented on Fig. \ref{SpinStiffness_N1212_Convergence}. The choice of 12 as the size of the rest grid dimensions is conditioned by keeping the resulting ${\cal D}$ stable (by means of relative error being less than 5\%, namely for 50$\times$8$\times$8 grid ${\cal D}$ = 884 meV~$\cdot$~\AA$^2$) while maximizing N$_{\mathrm{k}}$. As well as it was in real space consideration, thus converged ${\cal D}$ appears overestimated (844 meV~$\cdot$~\AA$^2$), of the same order of magnitude as experimental value. However, the fact of convergence itself indicates the capability of the proposed scheme to properly capture the itinerant contributions to magnetism in FeGe. By these means the reciprocal space consideration for ${\cal D}$ appears preferable.

\begin{table}[h]
\centering
\caption [Bset]{Calculated ${\cal D}$ (in meV~$\cdot$~\AA$^2$) using Eq. \ref{D_final_expression} with uniform \kmesh of the size N$_{\mathrm{k}}$$\times$N$_{\mathrm{k}}$$\times$N$_{\mathrm{k}}$.}
\label {D_with_uniform_KMesh}
\setlength{\tabcolsep}{5pt}
\begin {tabular}{c|cccc}
 \hline  \hline
 N$_{\mathrm{k}}$  &     8  &    12  &  16  &    20  \\
 ${\cal D}$        &  1858  &  1417  &  97  &  1588  \\
 \hline  \hline
\end {tabular}
\end {table} 

\section{Conclusion} \label{Sec:Conclusion} 

We performed the theoretical investigation of FeGe magnetic interactions, related to low-lying helical excitations of spin-wave dispersion spectrum. The high intensity of long-range pairwise exchange interactions, which reflect typical RKKY behaviour, prevents the real space magnetic picture from capturing the itinerant mechanisms exhaustively. Nevertheless, reproducing the experimental value of magnetic transition temperature in a reasonable accuracy keeps this approach applicable in mean-field framework.

The spin stiffness constant, being calculated in reciprocal space, manifests convergence as the size of \kmesh enhances. It confirms the general ability of distinct numerical scheme to model the observable collective magnetic excitations in real compounds with mixed type of magnetism. As a method for further enlarging the available real space exchange surrounding and improving the accuracy of reciprocal space calculations one could find useful the numerical interpolation \cite{kmesh_interpolation} of \kmesh in order to boost its size for fixed Hamiltonian. We also mention the prospects of its expansion by taking into account anisotropic Dzyaloshinskii-Moriya interaction in order to simulate the spin spiral state.

We emphasise that one needs the complementary theoretical approach to properly describe the observed helical state, with no prime focus on either local magnetic moments description, or itinerant picture. In present study we made a try to consolidate these two viewpoints by supplying a tool for delocalization assessment, which along with other microscopic models \cite{Blugel_MagneticModel, Arita, Dmitrienko_PRB, Dmitrienko_PRL} is expected to provide further insights in nature of intriguing topological traits, possessed by helimagnets.

\section{Acknowledgments}
We thank Vladimir Dmitrienko and Yaroslav Kvashnin for fruitful discussions. This work is supported by the grant program of Russian Foundation for Basic Research No. 16-32-00076.